\documentclass[trackchanges, twocolumn, twocolappendix]{aastex701}

\newcommand{\mytilde}{\raise.19ex\hbox{$\scriptstyle\sim$}}

\newcommand{\sersic}{S\'ersic }

\shorttitle{Intracluster Light of XLSSC~122}
\shortauthors{Joo et al.}

\received{}
\revised{}
\accepted{}

\submitjournal{ApJL}

\begin{document}

\title{Mature but Still Growing: JWST Detection of the Earliest Intracluster Light at $z\sim 2$}

\correspondingauthor{M. James Jee}
\email{gudwls4478@yonsei.ac.kr, mkjee@yonsei.ac.kr}

\author[orcid=0000-0001-9139-5455,gname=Hyungjin, sname=Joo]{Hyungjin Joo} 
\affiliation{
Department of Astronomy,
Yonsei University, 50 Yonsei-ro, Seoul 03722, Republic of Korea}
\email{gudwls4478@gmail.com}

\author[orcid=0000-0002-5751-3697,gname=Myungkook James, sname=Jee]{M. James Jee} 
\affiliation{
Department of Astronomy,
Yonsei University, 50 Yonsei-ro, Seoul 03722, Republic of Korea}
\affiliation{
Department of Physics,
University of California, Davis, One Shields Avenue, Davis, CA 95616, USA}
\email{mkjee@yonsei.ac.kr}

\author[orcid=0000-0002-4462-0709,gname=Kyle, sname=Finner]{Kyle Finner} 
\affiliation{
IPAC,
California Institute of Technology, 1200 E California Blvd, Pasadena, CA 91125, USA}
\email{kfinner@ipac.caltech.edu}

\author[orcid=0009-0009-4086-7665,gname=Zachary Pierce, sname=Scofield]{Zachary P. Scofield}
\affiliation{
Department of Astronomy,
Yonsei University, 50 Yonsei-ro, Seoul 03722, Republic of Korea}
\email{zpscofield@yonsei.ac.kr}  

\author[orcid=0000-0001-7148-6915,gname=Sangjun, sname=Cha]{Sangjun Cha} 
\affiliation{
Department of Astronomy, 
Yonsei University, 50 Yonsei-ro, Seoul 03722, Republic of Korea}
\email{sang6199@yonsei.ac.kr}

\author[orcid=0000-0003-2776-2761,gname=Jinhyub, sname=Kim]{Jinhyub Kim} 
\affiliation{
Department of Physics, 
University of Oxford, Denys Wilkinson Building, Keble Road, Oxford OX1 3RG, UK}
\email{jinhyub.kim@physics.ox.ac.uk}

\author[orcid=0000-0001-7583-0621]{Ranga-Ram Chary}
\affiliation{University of California, Los Angeles, CA 90095-1562, USA}
\email[]{rchary@ucla.edu}

\author[orcid=0000-0002-9382-9832]{Andreas Faisst}
\affiliation{IPAC, California Institute of Technology, 1200 E California Blvd., Pasadena, CA 91125, USA}
\email[]{afaisst@caltech.edu}

\author[orcid=0000-0003-1954-5046]{Bomee Lee}
\affiliation{Korea Astronomy and Space Science Institute, 776 Daedeokdae-ro, Yuseong-gu, Daejeon 34055, Korea}
\email[]{bomee@kasi.re.kr}

\begin{abstract}
We present a JWST analysis of intracluster light (ICL) in XLSSC~122 at $z=1.98$, currently the most distant known strong-lensing galaxy cluster with an evolved member population. 
Using deep JWST imaging complemented by HST data and careful control of systematics, we robustly detect diffuse emission extending to several hundred kpc from the brightest cluster galaxy (BCG) down to $\mytilde 29\ \mathrm{mag\ arcsec^{-2}}$. 
Multi-component PSF-convolved \sersic modeling separates the surface brightness profiles into three components: a BCG core, a BCG envelope, and an ICL component, with stable \sersic indices across wavelengths. 
Nearly flat color profiles indicate minimal radial variation in the stellar populations of the BCG envelope and the ICL.
The median ICL fraction measured across seven bands is $\mytilde 17\%$, demonstrating that the buildup of intracluster stars in massive halos was already well underway by $z \sim 2$. 
The ICL fraction peaks near $5000\ \text{\AA}$ in the rest-frame, resembling the behavior observed in dynamically active clusters. 
We also detect a southern excess of ICL relative to the best-fit \sersic model and quantify it using wavelet-based modeling, providing additional support that this system is dynamically active.
The BCG+ICL light distribution and strong-lensing mass map show strong morphological agreement within $\mytilde 100~\mathrm{kpc}$. 
These findings establish the ICL as an early-forming and dynamically informative component of massive halos.
\end{abstract}

\section{Introduction}
\label{introduction}
Intracluster light (ICL) was first reported by \citet{Zwicky1937} as faint emission between galaxies in the Coma cluster.
Since then, decades of studies have revealed that ICL constitutes a significant stellar component in galaxy clusters.
The ICL is composed of stars that are unbound from cluster galaxies and provides unique insights into galaxy evolution and the assembly history of massive structures.

The physical explanation for the formation of ICL remains in contention.
Some studies attribute most of the ICL to the formation of the brightest cluster galaxy (BCG) through mergers or violent relaxation \citep[e.g.,][]{Murante2007, Ko2018}, while others emphasize satellite tidal stripping \citep[e.g.,][]{Rudick2009, Contini2019} or pre-processing in galaxy groups prior to cluster accretion \citep[e.g.,][]{Rudick2006, Contini2014}.
Recent studies with cosmological simulations suggest that these channels are not mutually exclusive but instead form a continuous sequence. While multiple mechanisms contribute simultaneously to ICL buildup, the dominant driver varies by system based on its unique evolutionary history \citep{Joo2025}.

Observational and theoretical studies suggest that the ICL can provide valuable information on cluster assembly history. The ICL has been shown to be a sensitive tracer of the global dark matter distribution in clusters \citep[e.g.,][]{Jee2010, Montes2019, Yoo2024, Butler2025, Cha2025}.
Furthermore, \citet{Deason2021} and \citet{Gonzalez2021} emphasized the potential of the ICL as a diagnostic for identifying the splashback radius in clusters.
Beyond tracing dark matter, the ICL also provides valuable constraints on the dynamical state of clusters.
Recent studies suggest that the observed ICL fraction is a nearly constant function of wavelength in relaxed clusters, whereas unrelaxed systems tend to show a higher overall ICL fraction and a modest increase in the ICL fraction around $4800\ \text{\AA}$ \citep[e.g.,][]{Jimenez-Teja2018, Jimenez-Teja2021, deOliveria2022, Jimenez-Teja2024, deOliveria2025}. This enhancement in both the total ICL content and the blue spectral component likely reflects the contribution of recently stripped young stars during active merging events.
Simulations further demonstrate that the asymmetric spatial distribution of the ICL becomes more prominent when the host halo is not fully virialized \citep[e.g.,][]{Chun2024, Joo2025, Kimming2025, Jeon2025}.
Together, these properties make the ICL a powerful probe of cluster dynamical states and a unique test of hierarchical structure formation in the $\Lambda$CDM paradigm.

Because the ICL is accumulated over a wide range of epochs \citep[e.g.,][]{Contini2024, Brown2024, Joo2025, Jeon2025}, it preserves a cumulative record of past interactions and accretion events in clusters.
Unlike gas or galaxies, which can dynamically relax or dissipate on relatively short timescales, the diffuse stellar component retains imprints of the cluster’s assembly over several gigayears \citep{Rudick2009, Murante2007}. 
The spatial and structural properties of the ICL thus may provide a fossil record of hierarchical growth, offering a long-term view of how stellar material is redistributed within massive halos.

Probing the ICL at high redshift has long been challenging due to its extremely low surface brightness (SB).
Thanks to the Hubble Space Telescope (HST), observations have extended to $z \sim 1$–2 \citep[e.g.,][]{DeMaio2018, Joo2023}, but they are limited by depth and wavelength coverage.
For clusters at $z>2$, optical observations probe rest-frame wavelengths shortward of the Balmer break, where the ICL emission drops steeply.
Therefore, even at comparable SB limits, optical data generally lack the signal needed to trace the diffuse light reliably.
Unlike HST, which is limited to observations at wavelengths up to $1.7~\mu m$, JWST probes significantly longer wavelengths, making it a powerful tool for studying intracluster light at $z\gtrsim2$. The unprecedented infrared sensitivity of JWST now enables direct detections of diffuse stellar halos in more distant clusters.

In this context, XLSSU J021744.1-034536 ($z=1.98$, hereafter XLSSC~122) stands out as one of the most distant massive clusters observed to date \citep{Pierre2006, Pierre2016, Willis2013, Willis2020, Kim2025, Finner2025, Scofield2025b}.
With a look-back time of $\mytilde$10~Gyr, it provides a unique laboratory to test how early massive halos assembled their stellar and dark-matter components.
Although several galaxy overdensities at comparable redshifts have been reported \citep[e.g.,][]{Shi2024,Sun2024,Shimakawa2025}, these systems are often interpreted as being in a protocluster stage.
In contrast, previous studies have shown that XLSSC~122 is already a mature, gravitationally bound cluster, as evidenced by its quiescent galaxy population \citep{Willis2020, Noordeh2021} and mass measurements from X-ray and weak-lensing analyses \citep{Mantz2018, Kim2025}.
Recent SZ and weak-lensing analyses suggest that XLSSC~122 is dynamically active, consistent with ongoing merger activity \citep{Marrewijk2023, Scofield2025b}.

In this paper, we present an analysis of the ICL in XLSSC~122 using HST and JWST observations.
The combination of deep JWST imaging and strong-lensing (SL) constraints \citep{Finner2025} now enables the first detailed study of the cluster's extended diffuse stellar light and the relationship between its ICL and dark matter.
We find that this system exhibits a high ICL fraction, comparable to those of low-redshift clusters, and demonstrate that spatial and color variations offer insights into its dynamical state.
Section~\ref{sec:data} describes the observation and detailed data reduction.
Section~\ref{sec:result} presents the results of the BCG and ICL decomposition and measurements of the ICL fraction.
Section~\ref{sec:discussion} discusses the asymmetric ICL feature and the comparison with the SL mass map, and Section~\ref{sec:summary} summarizes the main results.
We report magnitudes in the AB system and adopt a flat $\Lambda$CDM cosmology with $H_{0}=70~\mathrm{km\ s^{-1}\ Mpc^{-1}}$ and $\Omega_{m}=0.3$ throughout this paper.
At the cluster redshift ($z=1.98$), this cosmology provides an angular scale of $1^{\prime\prime}=8.379~\mathrm{kpc}$.

\section{Data and Method} \label{sec:data}
\subsection{Data reduction}
In this work, we analyze HST and JWST observations of XLSSC~122. 
The HST data used in this analysis were obtained in three filters (F814W, F105W, and F140W; Program IDs: 15267 and 17172; PI: R. E. A. Canning).
The JWST imaging consists of four filters (F090W, F200W, F277W, and F356W; Program ID: 3950; PI: K. Finner).
These observation depths provide sufficient sensitivity to detect diffuse ICL at $z\sim2$, particularly with JWST, enabling reliable cross-filter comparison.

While initial reduction of the HST and JWST data followed standard calibration pipelines, we employed custom pipelines to address specific systematics critical to ICL analysis.
The HST observations were reduced using \texttt{AstroDrizzle} in the \texttt{DrizzlePac} package \citep{Hoffmann2021}.
The JWST data were processed with the standard JWST calibration pipeline \citep{Bushouse2022}.
Both the HST and JWST mosaics were resampled to a common pixel scale of $0.02^{\prime\prime}\ \mathrm{pixel^{-1}}$ using a square drizzle kernel.
Several systematics, such as wisp features, 1/f noise, sky gradients, and residual background, were corrected using customized post-processing routines for HST \citep{Joo2023} and JWST\footnote{\url{https://github.com/zpscofield/young-jwstpipe}} \citep{Scofield2025}, respectively.
For wisp features, we adopted Version 3 of Wisp Templates\footnote{\url{https://stsci.box.com/s/1bymvf1lkrqbdn9rnkluzqk30e8o2bne}} provided by the JWST User Documentation, which has $\mytilde$50\% less noise than Version 2\footnote{\url{https://www.stsci.edu/contents/news/jwst/2024/new-nircam-wisp-templates-are-now-available}}.
Further details on customized data reduction procedures are described in Appendix~\ref{appendix:reduction}.
These steps reduce instrumental systematics, which are particularly important for robust SB measurements of diffuse emission.

ICL analysis is sensitive to large-scale flux variations. Current in-flight calibrations have reduced large-scale flat-field systematic residuals (L-flats) to $\lesssim 1\%$ across the NIRCam detectors \citep{Sunnquist2024}. At this level of precision, the dominant systematics are no longer the flat-field response, but rather detector-level artifacts like $1/f$ noise and scattered light wisps, which we address through our custom  pipeline mentioned above.

\begin{figure*}
    \centering
    \includegraphics[width=0.98\textwidth]{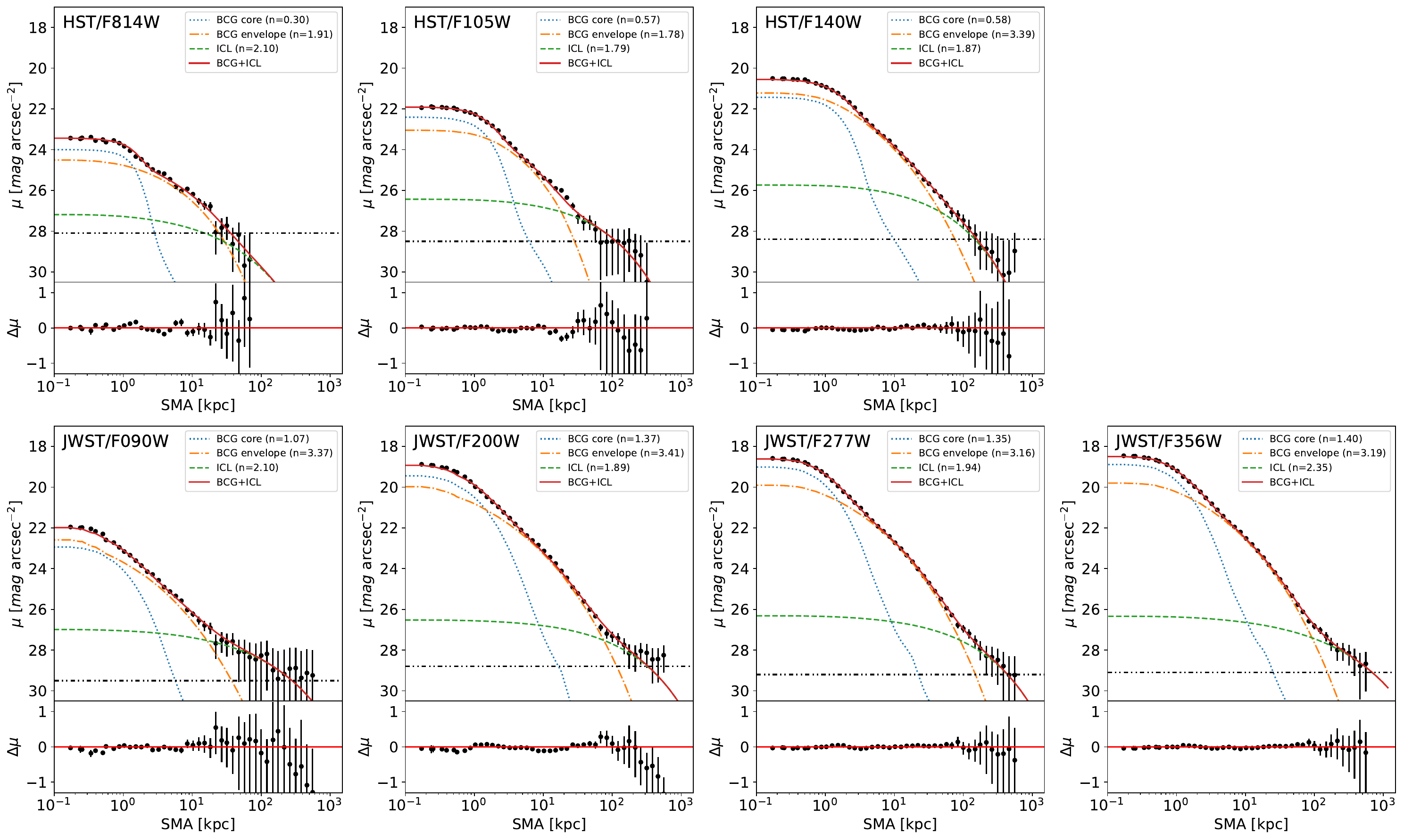}
    \caption{Multiband SB Profiles and Model Decomposition. 
    The top row displays HST observations (F814W, F105W, and F140W), while the bottom row presents JWST/NIRCam observations (F090W, F200W, F277W, and F356W).
    Black data points represent the observed median SB within elliptical bins, with error bars accounting for the quadrature sum of flux uncertainties and systematic sky error ($\sigma_{sky}$).
    Horizontal dot-dashed lines indicate the limiting SB ($\mu_{lim}$) for each respective band.
    The red solid line denotes the best-fit multi-component \sersic model convolved with the telescope PSF. The model is decomposed into three distinct stellar components: the BCG core (blue dotted line), the BCG envelope (orange dot-dashed line), and the diffuse ICL (green dashed line). The last components (green dashed line) are utilized for the ICL analysis in this work. Lower subpanels show the residuals ($\Delta\mu$) between the observed data and the total model.
    \label{fig:03_SBprofile}
    }
\end{figure*}

\begin{figure}
    \centering
    \includegraphics[width=0.97\linewidth]{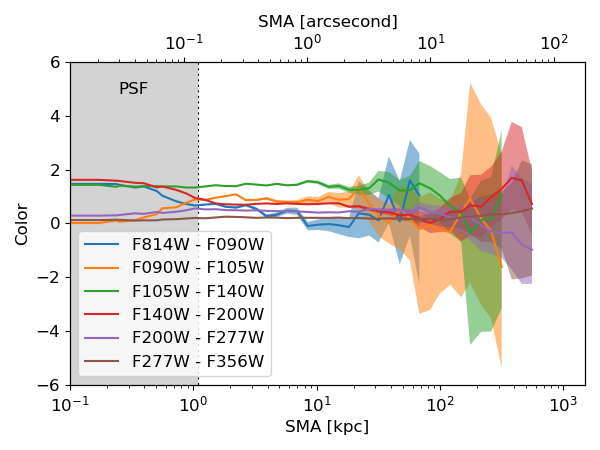}
    \caption{
    Radial color profiles of the BCG+ICL derived from six adjacent filter pairs.
    Shaded regions represent the 68\% confidence interval, incorporating both statistical flux uncertainties and systematic sky background errors ($\sigma_{sky}$).
    The inner gray region ($\lesssim 1$ kpc) indicates the area where differences in PSF widths between HST and JWST may induce artificial color gradients. 
    Beyond this scale, the color profiles remain remarkably flat within the error margins, suggesting a spatially uniform stellar population across the BCG envelope and ICL.
    A notable exception is the F814W$-$F090W profile, which exhibits variation at $3 \lesssim \rm SMA \lesssim 10$ kpc, potentially indicating radial differences such as dust attenuation within the central galaxy (see text).}
    \label{fig:color_profiles}
\end{figure}

\subsection{Surface Brightness Measurement} \label{subsec:sbmeasure}
Before measuring the SB profiles of the BCG and ICL, we masked out all discrete sources.
The mask map was constructed using the segmentation map generated by \texttt{SExtractor}.
To further minimize the influence of extended sources, the mask map was expanded using a binary dilation method \citep{matheron1988}.
The expansion width ($w$) for each source was scaled individually using the following relation:
\begin{equation}
    w = c_{e} \times r_{m},
\label{eq:expansion_width}
\end{equation}
where $c_{e}$ is the expansion coefficient, and $r_{m}$ is the radius derived from the area of the original segmentation mask, defined as $r_{m} = \sqrt {Area / \pi}$.
While \cite{Joo2023} used the half-light radius of each source, this approach was replaced by $r_{m}$ in this work to better handle bright stellar sources, which are not well represented by half-light radii.
We adopted $c_{e} = 1.2$ to minimize contamination from other light sources while avoiding excessive masking of the ICL.
The BCG itself was excluded from the masking process to preserve its flux profile.

After applying the masks, we defined elliptical bins centered on the BCG that matched the elongated BCG + ICL morphology.
The global ellipticity and position angle, accounting for both the BCG and ICL, were measured using \texttt{AUTOPROF} \citep{Stone2021} from the detection image constructed by combining the four JWST filters (F090W, F200W, F277W, and F356W).
We obtained a global ellipticity ($1-b/a$) of $0.458\pm0.011$ and a position angle of $102.3^\circ\pm0.288^\circ$, measured counterclockwise from the east.
Using a background region outside the $1~\rm{Mpc}$ ellipse, we estimated the SB limits, $\mu_{lim}$ and the sky estimation uncertainties, $\sigma_{sky}$ for each band (Appendix~\ref{subappendix:sky}).
The elliptical radial bins were defined with semi-major axes increasing logarithmically.
We estimated the median SB level within each bin and its associated measurement error.
The errors were estimated by adding the flux uncertainties and $\sigma_{sky}$ within each radial bin in quadrature.
The radial SB profiles are presented in Figure~\ref{fig:03_SBprofile} for all seven bands.

Color profiles are a useful tool for comparing the stellar populations of the BCG and the ICL. 
We constructed color profiles from the SB profiles of adjacent filter pairs so as to trace the local slope of the underlying spectral energy distribution (SED) more directly.
Figure~\ref{fig:color_profiles} shows color profiles constructed from the six adjacent filter pairs: F814W-F090W, F090W-F105W, F105W-F140W, F140W-F200W, F200W-F277W, and F277W-F356W, following the wavelength ordered sequence of the filters.
The uncertainty in each color profile was calculated by adding the uncertainties in the two SB profiles in quadrature.
The discussion of the results is presented in \textsection~\ref{subsec:decomposition}.

\section{Result} \label{sec:result}
\subsection{Radial Profiles of BCG and ICL} \label{subsec:decomposition}
As shown in Figure~\ref{fig:03_SBprofile}, the ellipsoidal profiles reveal the existence of diffuse stellar light beyond a few hundred kpc from the BCG in most bands, thereby enabling direct constraints on the spatial extent of the diffuse stellar component at $z\sim 2$.
In F814W, which probes rest-frame wavelengths blueward of the Balmer break ($\mytilde 3645\text{\AA}$), the diffuse emission appears weaker than in the other bands. 
While F090W and F814W cover a similar spectral range, F090W’s higher throughput at longer wavelengths, combined with NIRCam’s high sensitivity, extends our detection of the ICL to greater radial distances from the BCG, despite its shorter exposure time.
In both HST and JWST images, the diffuse signal strengthens toward longer wavelengths, consistent with emission from evolved stars and the scarcity of young, blue stars in the ICL at $z \sim 2$.

To quantify the contributions of the ICL to these profiles, we decomposed the SB profiles into multiple components as described below.
We assumed that each component follows a \sersic profile \citep{Sersic1963} and modeled the overall profile as a sum of \sersic functions, as shown in Equation~\ref{equation:multi_sersic}.
\begin{equation}
\mu(R) = \\ \mathrm{PSF}(R) * \sum_{i}^{N} I_{e,i} \exp\left[-k\left(\left(\frac{R}{R_{e,i}}\right)^{{1}/{n_i}}-1\right)\right]
\label{equation:multi_sersic}
\end{equation}
Here, the model is convolved with the point spread function (PSF) to account for both the blurring effect and the diffraction spikes. 
We used \texttt{TinyTim} \citep{Krist2011} to simulate the PSF of HST and \texttt{STPSF} \citep{Perrin2014} for JWST.

Before decomposing the BCG and ICL, we determined the number of required \sersic components.
First, we calculated the Bayes factor ($K$) by comparing the evidence value of two different models: two-component and three-component \sersic models.
$K_{32}$ is defined as in Equation \ref{eq:bayes}:
\begin{equation}
    K_{32}
    =
    \frac{\Pr(D|M_3)}{\Pr(D|M_2)}
    \label{eq:bayes}
\end{equation}
where $D$ denotes the observed data, and $M_2$ and $M_3$ correspond to two- and three-component models, respectively.
We used \texttt{PyMultiNest} \citep{Buchner2014} to obtain the evidence for each model.
The resulting $\log_{10}(K_{32})$ values are 0.58 (F814W), 1.22 (F105W), 1.16 (F140W), 1.10 (F090W), 1.02 (F200W), 1.24 (F277W), and 1.05 (F356W).
According to Jeffreys' scale \citep{Jeffreys1998}, the three-component model is ``substantially'' preferred to the two-component model for F814W profile, even though the profile reaches its limiting depth at $\mytilde 50$ kpc.
For the other bands, the three-component models are ``strongly'' favored. with the Bayes factors greater than 10.
Thus, we adopted three \sersic components to model the BCG and ICL profiles.
Figure~\ref{fig:03_SBprofile} presents the result of three-component modeling with \texttt{PyMultiNest}. 

To further validate the requirement of three components, we examined additional independent criteria proposed in previous studies.
\citet{Gonzalez2021} defined the BCG-ICL transition as the location of a dip in the derivative of the SB profile, typically found on scales of several tens of kpc, consistent with both simulations \citep[e.g.,][]{Contini2022} and observations \citep[e.g.,][]{Montes2017}.
Building on this approach, \citet{Joo2023} proposed using the number of such dips as an indicator of the number of \sersic components, adopting one more component than the number of detected dips.
Similarly, \citet{Zhang2019} found that three \sersic components were required to reproduce stacked DES cluster profiles.
In our data, we identified two dips in the SB profiles for six filters, while only a single dip is detected in F814W.
The latter is likely due to the shallower $\mu_{lim}$ of F814W, which reached only $\mytilde50$ kpc and limits the detectability of the diffuse ICL component.
The consistency between the dip-based expectations and Bayes factor selection therefore provides an independent check on our adopted three-component decomposition.

Figure~\ref{fig:color_profiles} presents the color profiles of the BCG and the ICL across six filter combinations.
Beyond $\mytilde10~\rm{kpc}$, all six colors exhibit nearly flat radial gradients, indicating little change in the stellar populations with distance from the center.
At $\rm{SMA} \lesssim 3~\rm{kpc}$, radial color trends appear in the F814W-F090W, F090W-F105W, and F140W-F200W colors.
These colors are constructed from filters observed with different telescopes and therefore have substantially different PSF widths (e.g., $\mytilde0.08^{\prime\prime}$ for F814W and $\mytilde0.03^{\prime\prime}$ for F090W), suggesting that the trends are likely to be driven by PSF mismatch rather than stellar population differences.
A noticeable variation is detected only in F814W-F090W at $3\lesssim \rm{SMA}\lesssim 10~\rm{kpc}$, corresponding to the BCG-dominated region.
Because this color corresponds to the rest-frame near-ultraviolet, the observed behavior likely indicates intrinsic properties of the central galaxy, such as dust attenuation.
Further analysis of the BCG is required to identify the physical driver of this variation.

Meanwhile, the \sersic indices are broadly consistent across the seven bands: the BCG cores are approximately exponential ($n\sim1$), the BCG envelopes show steeper profiles with $n\sim3$, and the ICL components remain relatively diffuse with $n\sim2$.
Although this cluster is located at $z=1.98$, the profile decomposition reveals a structure similar to the stacked profiles at $z\sim0.25$ reported by \citet{Zhang2019}.
Taken together, these results indicate that the BCG and ICL in XLSSC~122 have already settled into a smooth, extended structure by $z\sim2$. 
If we consider the mass of this halo ($M_{200,c} = 2.6\pm1.1\times 10^{14} M_{\odot}$ from SL modeling; \citealt{Finner2025}), the presence of a mature structure is not unexpected \citep[e.g.,][]{Contini2024, Joo2025, Kimming2025}.
These findings establish XLSSC~122 as the earliest cluster in which the structural maturity of the ICL can be directly characterized. 

\begin{figure*}
    \centering
    \includegraphics[width=0.85\textwidth]{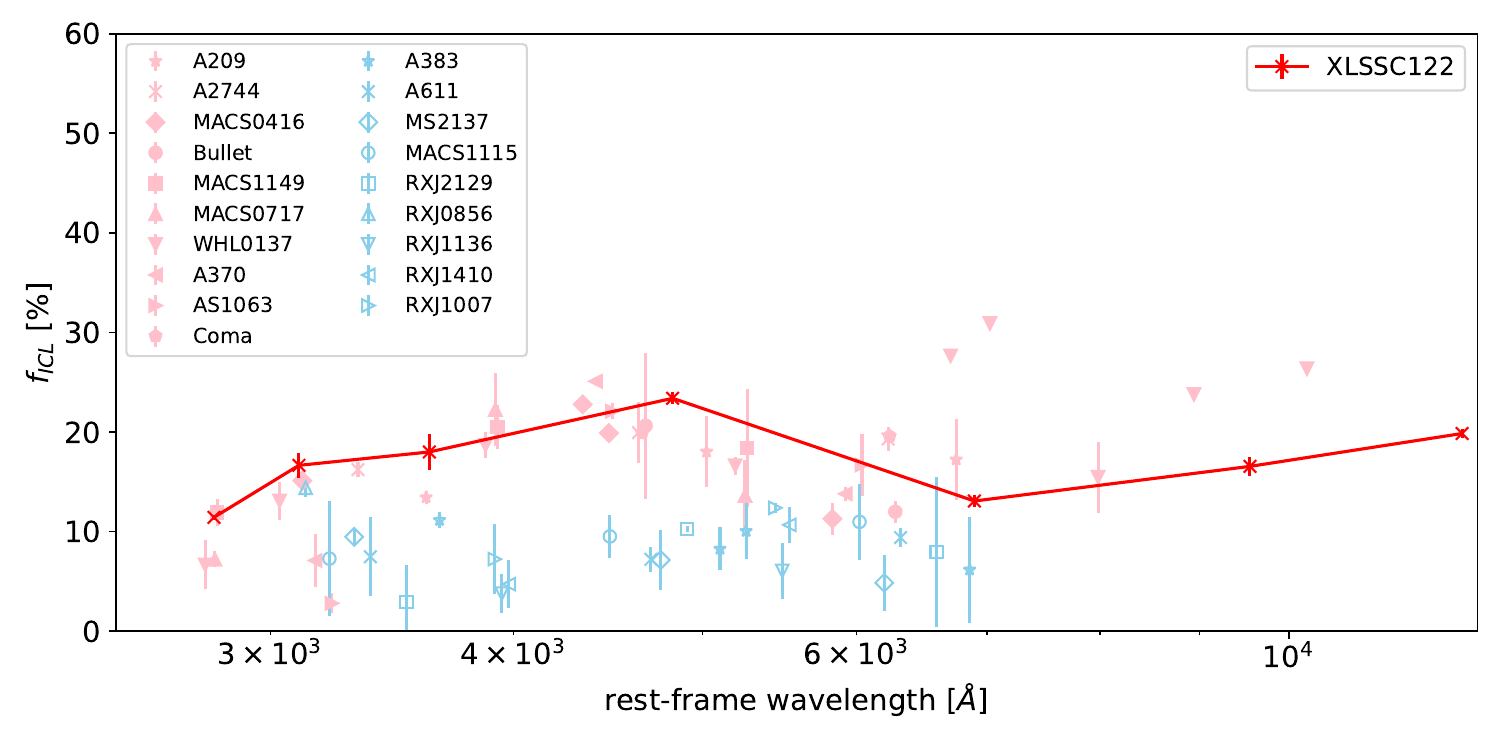}
    \caption{
    ICL fraction ($f_{ICL}$) as a function of rest-frame wavelength. 
    The red crosses represent the measurements for XLSSC~122 ($z=1.98$) across the seven observed bands.
    For comparison, pink (sky blue) points indicate $f_{ICL}$ in dynamically active (relaxed) clusters measured by \citet{Jimenez-Teja2018, Jimenez-Teja2021, deOliveria2022, Jimenez-Teja2024, deOliveria2025}.
    The $f_{ICL}$ values for XLSSC~122 exhibit a characteristic local maximum near 4800 $\text{\AA}$, a pattern that closely resembles the trends found in dynamically active or merging systems. 
    }
    \label{fig:04_frac_wave}
\end{figure*}

\subsection{ICL Fraction} \label{subsec:icl_fraction}

Based on the \sersic decomposition, we next measured the ICL flux fraction at each wavelength.
We define the ICL fraction as follows:
\begin{equation}
    f_{ICL} = \frac{F_{ICL}}{F_{ICL} + F_{galaxy} }
    \label{equation:icl_fraction}
\end{equation}
where $F_{galaxy}$ denotes the flux from XLSSC~122 member galaxies, including the BCG, selected via either spectroscopic or photometric criteria (see Appendix~\ref{appendix:member}).
For each filter, $F_{galaxy}$ and $F_{ICL}$ are estimated within the specific radius where the SB profile intersects the limiting magnitude ($\mu_{\text{lim}}$).
This criterion ensures that the ICL is measured only in regions with sufficient signal-to-noise ratio, thereby avoiding biases from extrapolated signals.
The ICL fractions measured out to the limiting radius in each band are $11.4\pm0.45\%$ within 21.9 kpc (F814W), $18.0\pm2.98\%$ within 68.5 kpc (F105W), $23.4\pm1.04\%$ within 177.6 kpc (F140W), $16.6\pm2.13\%$ within 267.2 kpc (F090W), $13.1\pm0.94\%$ within 319.4 kpc (F200W), $16.6\pm1.65\%$ within 460.7 kpc (F277W), and $19.8\pm0.78\%$ within 557.4 kpc (F356W).
The average fraction in the seven bands is $\mytilde 17\%$.

Previous studies based on HST observations have provided constraints on the ICL fraction at $z \sim 2$.
\citet{Joo2023} reported an ICL fraction of approximately 10\% in JKCS041 at $z = 1.9$.
\citet{Werner2023} further showed that in systems at $z \sim 2$, such as CARLA J1018+053, the ICL component appears to be strongly concentrated toward the central regions.
However, due to the limited wavelength coverage and sensitivity of HST, ICL measurements in systems at $z \gtrsim 2$ are restricted to small apertures near the BCGs.
This study demonstrates that such limitations can be alleviated using JWST, enabling robust quantification of the ICL fraction at higher redshifts.
Additionally, \citet{Coogan2023} reported an ICL fraction of approximately 10\% in a proto-cluster at $z=1.85$.
While this value is slightly lower than measured in this work, it remains consistent with the mass-dependent trend of ICL fraction suggested by \citet{Mayes2025} and \citet{Joo2025} from groups to clusters.
This study indicates that the ICL fraction measured in massive halos at $z \sim 2$ is consistent with previous estimates based on HST observations.
These results indicate that the build-up of the ICL fraction in massive galaxy clusters was already well established by $z \sim 2$, consistent with early assembly of diffuse stellar components.

\section{Discussion} \label{sec:discussion}

\subsection{Wavelength Dependence of ICL Fraction} \label{subsec:iclfraction_dependence}
Recent studies suggest that both the total amount and the wavelength dependence of the ICL fraction reflect the dynamical state of galaxy clusters. On average, merging or dynamically active systems exhibit higher ICL fractions than their relaxed counterparts. Furthermore, these dynamically disturbed clusters tend to show a characteristic rise in the ICL fraction near rest-frame 4800 \AA, whereas relaxed systems display a comparatively flat trend across wavelengths \citep{Jimenez-Teja2018, Jimenez-Teja2021, deOliveria2022, Jimenez-Teja2024, deOliveria2025}.

Multiwavelength studies have suggested that XLSSC~122 might be an active system \citep[e.g.,][]{Marrewijk2023, Scofield2025b}.
To investigate whether the ICL of XLSSC~122 follows the aforementioned trend, we studied the rest-frame wavelength dependence of the ICL fraction (Figure~\ref{fig:04_frac_wave}).
The fraction remains relatively low below $3000 \text{\AA}$, increases toward a local maximum near $4800 \text{\AA}$, decreases up to $\approx 7000 \text{\AA}$, and increases again at longer wavelengths.
The ICL fraction of XLSSC~122 shows a wavelength-dependent pattern broadly consistent with that observed in merging or dynamically disturbed clusters at lower redshifts.
This suggests that this wavelength-based probe of dynamical state might hold even at $z\sim2$.

According to \citet{Jimenez-Teja2021}, the enhanced ICL fraction around $4800 \text{\AA}$ likely reflects the contribution from recently stripped young stars, whose bluer spectral energy distributions raise the measured fraction in dynamically active clusters.
The observed structure in XLSSC~122 aligns with this physical interpretation, suggesting that tidal stripping and galaxy interactions are ongoing.

\subsection{ICL Excess Southward of the BCG} \label{subsec:assymicl}
\begin{figure}
    \centering
    \includegraphics[width = 0.98\columnwidth]{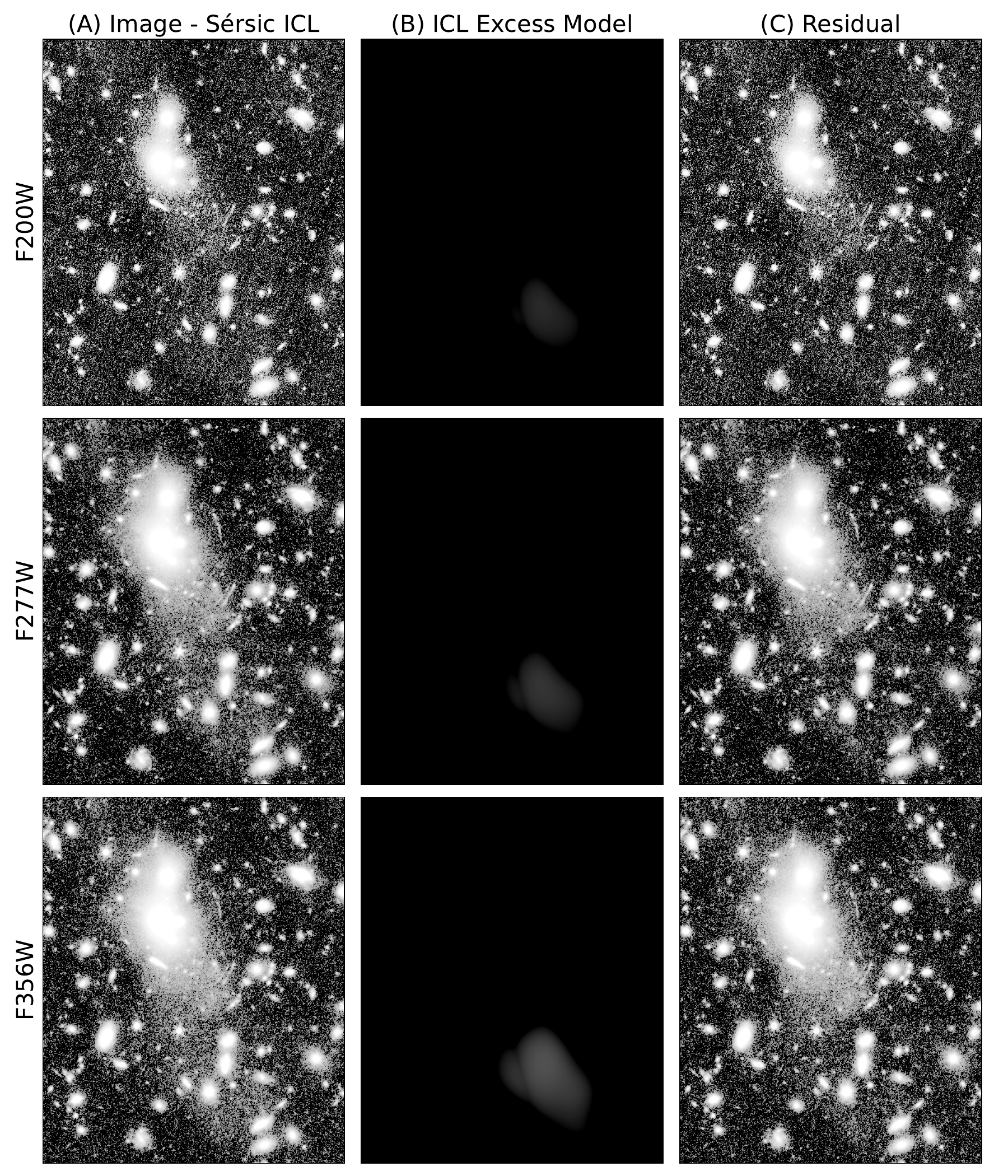}
    \caption{
    Localization of non-\sersic diffuse emission in XLSSC~122. 
    The left column presents JWST/NIRCam images (F200W, F277W, and F356W) following the subtraction of the best-fit 2D \sersic ICL (diffuse) component, revealing a prominent residual signal to the south. The middle column displays the isolated model of this additional diffuse component as detected by the wavelet-based algorithm {\tt DAWIS}. The right column shows the final residual maps after subtracting both the global \sersic and local {\tt DAWIS} models.
    This southern feature represents a significant excess of diffuse emission extending approximately 100 kpc from the BCG. Its spatial alignment with the overdensity of member galaxies and independent multi-wavelength asymmetries (X-ray, radio, and SZ) strongly suggests an origin in tidally stripped stars resulting from ongoing or recent dynamical interactions within the cluster.
    }
    \label{fig:05_additional_ICL}
\end{figure}

Visual inspection of our JWST imaging data reveals an excess of diffuse emission near the member galaxies southward of the BCG.
This feature is located at a projected distance of $\gtrsim100~\mathrm{kpc}$ from the BCG, and is still visible even after the subtraction of our \sersic ICL model (i.e., outermost component).
This signal is most significant in the F200W, F277W, and F356W filters, implying that its emission is dominated by a stellar population similar to that of the BCG and the ICL component in the \sersic model.
We show the image of XLSSC~122 after subtracting the 2D \sersic model of ICL in the left column of Figure~\ref{fig:05_additional_ICL}.
This asymmetric excess departs from the symmetry expected from a relaxed ICL distribution. 
This offers observational support for the ICL substructure associated with dynamical activity in the central region.

To measure the flux of this southern ICL excess, we modeled the excess ICL with \texttt{DAWIS} \citep{Ellien2021}, an algorithm designed to detect the intracluster light with wavelets.
The middle column of Figure~\ref{fig:05_additional_ICL} shows the model generated by \texttt{DAWIS}.
The modeled ICL is distributed across the member galaxies south of BCG. Nevertheless, identifying specific host galaxies of this ICL component remains a challenge.
Instead, the morphology of the excess broadly follows the southward elongation seen in the smoothed member density distribution (see the left panel of Figure~\ref{fig:01_colormap}).
And we find no clear single massive galaxy counterpart at its location.
These properties favor an origin in diffuse stellar debris produced by tidal stripping during recent galaxy interactions. 
Similar asymmetric ICL substructures are commonly found in simulations of dynamically unrelaxed clusters \citep[e.g.,][]{Joo2025, Jeon2025}.

The results in \citet{Joo2025} suggest that around 10\% of the ICL in $z\sim2$ or $M_{200} \geq 10^{14} M_{\odot}$ clusters is contributed by stars stripped from galaxies.
In this study, the southern component accounts for $\mytilde4\%$ of the total ICL flux, defined as the sum of the \sersic model ICL component and the southern ICL excess, in the three filters where it is most prominent.
While this value, if entirely originating from stripping, is lower than the average stripping contribution predicted by theoretical models, it is not unreasonably small when considering the expected cluster-to-cluster variation.
Rather than indicating a fundamental inconsistency, this difference likely reflects the fact that the observed excess traces only a subset of the total stripped component.
In particular, our modeling may miss more smoothly distributed or dynamically mixed stripped stars, and thus the detected excess should not be interpreted as representing all stripping events.
Nevertheless, the presence of this localized ICL feature provides evidence for ongoing or recent interactions among cluster member galaxies.
This asymmetric excess provides additional observational support that XLSSC~122 is dynamically disturbed, as inferred directly from its ICL.

In addition to the diffuse optical excess, independent multi-wavelength datasets show consistent signatures of asymmetry along the same axis.
X-ray imaging reveals an elongated morphology aligned with the asymmetric ICL \citep{Mantz2014}, and radio observations \citep{Hale2025,Jarvis2016} show a peak located near this overdense region \citep{Scofield2025b}.
Furthermore, the signal-to-noise peak of the Sunyaev-Zel'dovich effect is observed near the location of the asymmetric ICL feature \citep{Marrewijk2023}, suggesting the presence of bulk gas motions or ongoing interactions.
These multi-wavelength indications collectively support the interpretation that the southern diffuse component originates from recent or ongoing dynamical activity within the cluster.
For a more detailed comparison, including the X-ray, radio, and SZ analyses, readers are referred to \citet{Scofield2025b}.

\begin{figure*}
    \centering
    \includegraphics[width = \textwidth]{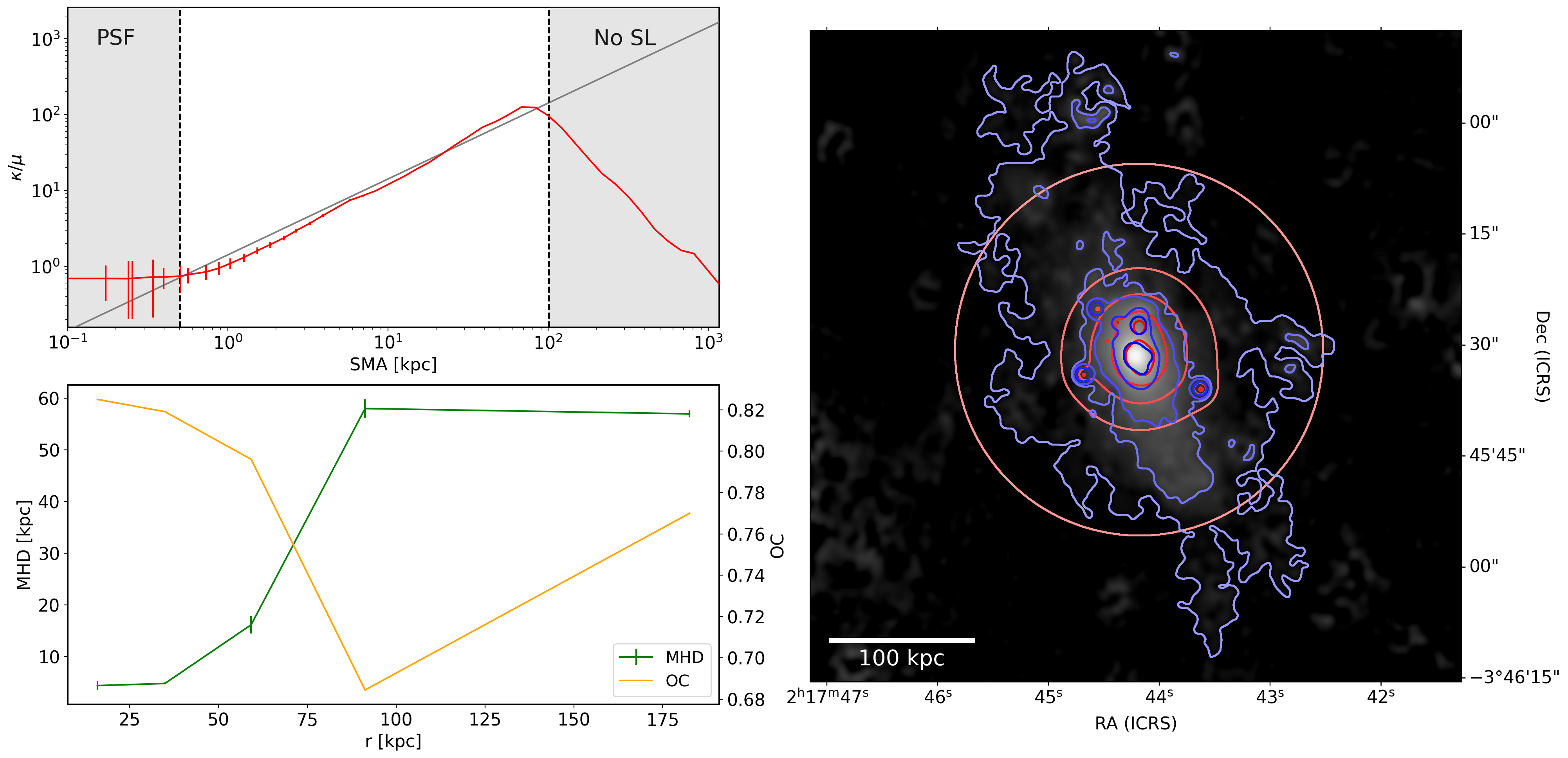}
    \caption{Comparison between the mass distribution and diffuse stellar light in XLSSC~122.
    Top left:
    The radial profile of the mass-to-light ratio ($\kappa/\mu$) is derived 
    from the SL mass map and the F356W SB profile. The ratio follows a linear relation ($\kappa/\mu \propto r$) up to $r \sim100$ kpc, the regime where the SL model constraints are most robust.  
    Right:
    A two-dimensional comparison between the SL mass distribution (red contours) and the smoothed stellar light distribution of the combined BCG + ICL + member galaxies (blue contours and gray scales) is presented.
    Bottom left: The orange curve illustrates the OC between the light and mass contours, while the green curve displays the MHD.
    Both metrics consistently indicate that the overall morphological similarity between mass and diffuse light components remains high up to  $r\sim 180$~kpc.
    The slight decrease in similarly beyond $r\sim60$~kpc is attributed to the presence of the southern ICL excess and the lack of the SL  constraints.}    
    \label{fig:06_comp_icl_mass}
\end{figure*}

\subsection{Comparison with Strong Lensing Mass}
To assess the connection between the ICL and the cluster potential, we compared the mass distribution with the BCG + ICL light distribution.
This offers a direct observational test of whether the BCG + ICL light distribution traces the SL mass distribution at $z\sim2$.
Recently, \citet{Yoo2024} and \citet{Butler2025} found that the dark matter to BCG+ICL density ratio profile is an approximately linear function of radius.
This relation may be the result of differences between the underlying physics of dark matter assembly and baryonic evolution.
In the CDM paradigm, baryons undergo radiative cooling and energy dissipation to form highly concentrated stellar components at galaxy scales, whereas dark matter remains as an extended, collisionless halo.
Within the cluster potential, the stellar and dark matter components of a galaxy are redistributed differently because of their distinct internal binding energy distributions.
This leads to a BCG+ICL distribution that is more centrally concentrated than the dark matter profile.

The top left panel of Figure~\ref{fig:06_comp_icl_mass} shows the lensing mass profile $\kappa(r)$ divided by the BCG+ICL SB profile $\mu(r)$. 
This profile is constructed using the F356W band, which provides the strongest BCG+ICL signal and the deepest radial extent. 
The innermost region of the SB profile ($\mytilde 0.6~\mathrm{kpc}$) is artificially flattened by the PSF. 
Beyond this radius, the ratio $\kappa / \mu$ scales approximately linearly with radius ($\kappa /\mu \propto \sim r$), consistent with the $r$ dependence reported in previous studies. 
This scaling breaks down beyond $\mytilde 100~\mathrm{kpc}$, where the SL constraints become sparse due to the rapid decline in the number of multiple images. 
This radius also coincides with the southern ICL excess identified in \textsection~\ref{subsec:assymicl}, further complicating the comparison between the mass and light distributions.

We also compared the 2D morphological similarity between the BCG + ICL and the SL mass distribution from \citet{Finner2025}.
The BCG and ICL are known to follow the cluster dark matter distribution \citep{Montes2019, Diego2023, Cha2025}.
To make this comparison, we constructed a 2D light distribution by first masking all detected sources in the same manner as in \textsection~\ref{subsec:sbmeasure}. 
Then, we filled the masked area with our BCG+ICL model from \textsection~\ref{subsec:decomposition} and applied a Gaussian smoothing kernel with $\sigma = 25$ pixels to enhance the large scale flux distribution.
The background of the right panel in Figure~\ref{fig:06_comp_icl_mass} shows the BCG, ICL, and galaxy flux distribution used for comparison with the mass.

To quantify the above morphological similarities, we use the weighted overlap coefficient \citep[WOC]{Yoo2022}. WOC evaluates the similarity of two contour sets by measuring the area of the overlapping regions, yielding higher values for greater similarities.
The two contour sets that we used are displayed in the right panel of Figure~\ref{fig:06_comp_icl_mass}.
The red contours are from the SL mass map and the blue contours are from BCG + ICL + galaxies.
The orange line in the lower left panel of Figure~\ref{fig:06_comp_icl_mass} shows the overlap coefficient (OC) at each radius\footnote{WOC is a weighted average of  OC values at all radii.}.
The OC value at the central region ($r\sim20~\mathrm{kpc}$) was about $0.82$, indicating that the BCG+ICL distribution closely traces the mass in the inner halo.
Farther from the center, the OC value slightly decreases, reaching 0.78 at $r\sim60~\mathrm{kpc}$.
This decline is expected given the ellipticity mismatch between the mass and ICL distributions in this region.
Nevertheless, the level of overall agreement  at $r\lesssim60~\mathrm{kpc}$ remains high.
The similarity drops more steeply beyond $r\gtrsim60~\mathrm{kpc}$, reaching OC $\sim 0.69$ at $r\sim 100 ~\rm{kpc}$.
At this radius, we find an excess of ICL on the southern side of the BCG (\textsection~\ref{subsec:assymicl}), which  is not present in the mass map.
Beyond this southern excess region, the agreement increases again ($\rm{OC}\sim0.77$ at $r\sim180 ~\rm{kpc}$).
The WOC value (i.e., weighted average of all OC values) is 0.813, indicating a high level of agreement between the overall light and mass distributions.

In addition to WOC, we also employed the Modified Hausdorff Distance \citep[MHD]{Dubuisson1994} to quantify the comparison, where lower values indicate higher similarities.
We used the same set of contours adopted for the WOC calculation.
The green line in the lower left panel of Figure~\ref{fig:06_comp_icl_mass} shows the resulting MHD values.
The mean MHD value is $\mytilde 28.1~\rm{kpc}$.
The MHD result is highly consistent with the OC trend, showing high similarity in the inner region and low similarity beyond $\mytilde 100~\mathrm{kpc}$.

Overall, OC and MHD show that the BCG+ICL closely follows the SL mass map at  $r \lesssim 100~\rm{kpc}$, providing direct observational evidence that the stellar distribution traces the cluster potential at $z\sim2$. 
The reduced agreement at larger radii likely reflects the sparsity of multiple image constraints.
In this context, weak lensing extends the comparison to larger scales, showing improved agreement around $r \gtrsim 100~\rm{kpc}$ \citep{Scofield2025b}. 
Moreover, the linear regime of $\kappa/\mu$ extends from $\mytilde 100~\rm{kpc}$ in the SL comparison to $\mytilde 200~\rm{kpc}$ when $\kappa(r)$ is taken from the weak lensing reconstruction.

\section{Summary} \label{sec:summary}
In this work, we analyzed the ICL in XLSSC~122 at $z=1.98$ and detected diffuse emission extending to several hundred kpc from the BCG.
The diffuse signal is strong in the rest-frame optical but fainter in the ultraviolet, consistent with an evolved stellar population dominating the diffuse components.

The SB profiles require three components: a compact BCG core, an extended BCG envelope, and an ICL component.
This decomposition is stable across all filters, with broadly similar \sersic indices.
The overall color profiles are flat, implying minimal radial variation in the dominant stellar populations of the BCG envelope and the ICL.

The ICL fraction has a mean value of $16.8\%$.
The fraction peaks at $\mytilde 4800~\text{\AA}$ in the rest-frame. This wavelength dependence resembles the trend reported for dynamically active clusters.

We also detected a significant ICL excess $\gtrsim 100~\rm{kpc}$ south of the BCG in F200W, F277W, and F356W.
Its location coincides with the southern overdensity in the member galaxy distribution and is consistent with the asymmetry seen in X-ray, SZ, and radio.
This excess emission contributes $\mytilde4\%$ of the total ICL flux and may be attributed to tidal stripping from galaxy interactions.

Within $\mytilde 100 ~\rm{kpc}$, the BCG+ICL distribution follows the cluster mass ($\rm{WOC}\sim0.813,~ \rm{MHD}\sim 28~\rm{kpc}$), and the $\kappa / \mu $ profile linearly increases with radius.
This indicates that the ICL can trace the dark matter distribution even in the $z\sim2$ regime.
Taken together, our results demonstrate that a mature yet still growing ICL component was already present in a $ M_{200}>10^{14}\ M_{\odot}$ halo at $z\sim2$. 

\begin{acknowledgments}
This work is based on observations made with the NASA/ESA/CSA JWST and downloaded from the Mikulski Archive for Space Telescopes (MAST) at the Space Telescope Science Institute (STScI), which is operated by the Association of Universities for Research in Astronomy, Inc., under NASA contract NAS 5-03127 for JWST. These observations are associated with program \#3950.
Support for program \#3950 was provided by NASA through a grant from the Space Telescope Science Institute, which is operated by the Association of Universities for Research in Astronomy, Inc., under NASA contract NAS 5-03127.
The JWST data presented in this article were obtained from the Mikulski Archive for Space Telescopes (MAST) at the Space Telescope Science Institute. The specific observations analyzed can be accessed via \dataset[doi: 10.17909/hemh-ca02]{https://doi.org/10.17909/hemh-ca02}.
H. Joo acknowledges support from the International Joint Research Grant of Yonsei Graduate School.
MJJ acknowledges support for the current research from the National Research Foundation (NRF) of Korea under the programs 2022R1A2C1003130 and RS-2023-00219959.
B. Lee is supported by the NRF grant funded by the Korea government (MSIT), 2022R1C1C1008695

The HST and JWST data presented in the Letter were obtained from MAST at STScI.
The observations  can be accessed via DOI: \url{doi:10.17909/16ah-g805}.
\end{acknowledgments}

\begin{contribution}
H. Joo led the ICL analysis, reduced HST/JWST data, estimated photometric redshifts, and wrote the paper.
M. James Jee supervised the project and contributed to manuscript writing.
Z. Scofield assisted with the JWST/NIRCam data reduction and participated in the interpretation of the results.
J. Kim assisted with the HST data reduction.
K. Finner and S. Cha provided the strong-lensing results and participated in the interpretation.
\end{contribution}

\bibliography{reference}
\bibliographystyle{aasjournal}

\begin{appendix}

\section{Customized Data Reduction} \label{appendix:reduction}
\begin{figure*}
    \centering
    \includegraphics[width=0.9\textwidth]{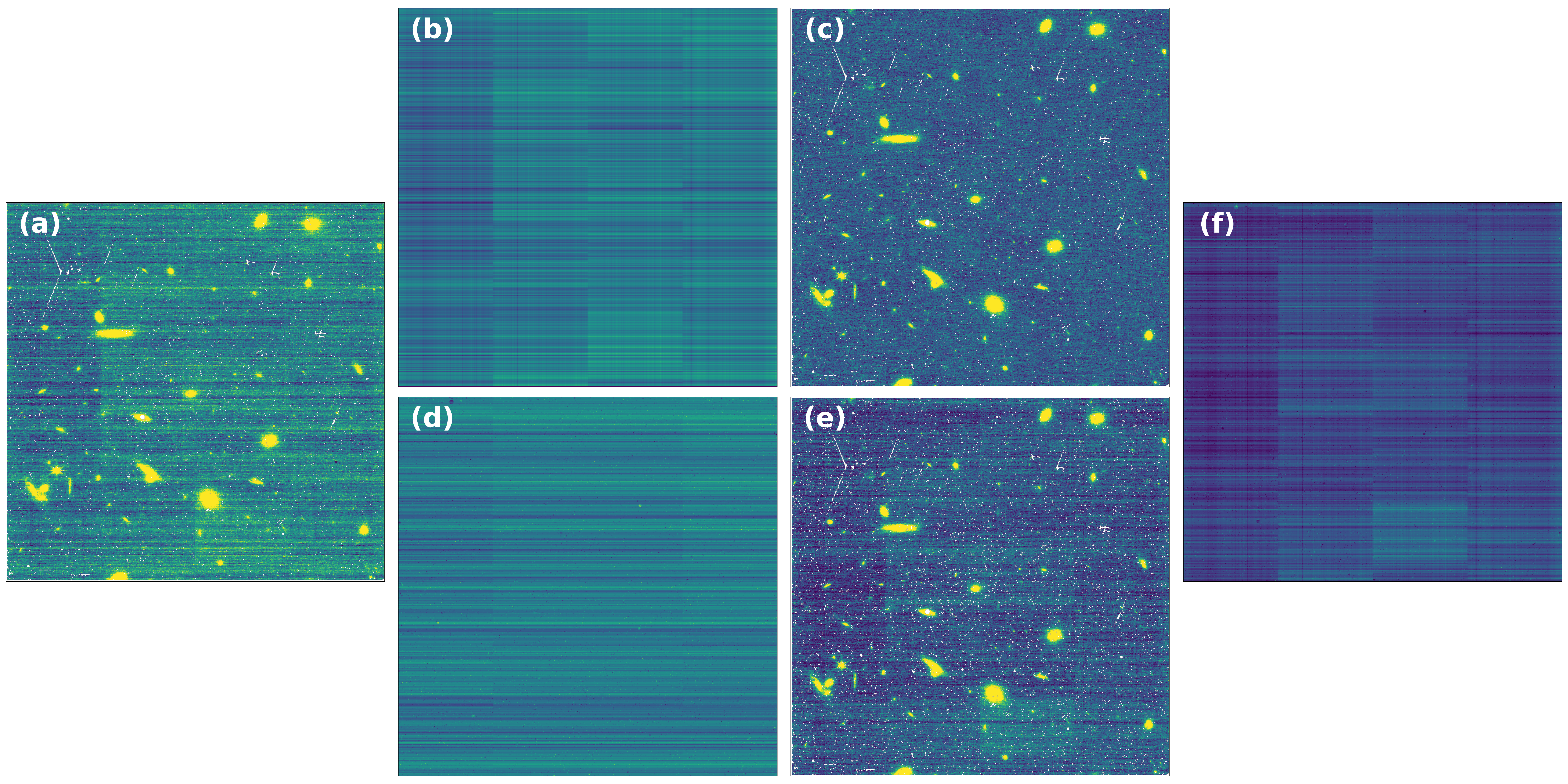}
    \caption{Example of modeling the 1/f noise.
    Panel (a) shows the original image.
    Panels (b) and (c) present the modeled 1/f noise and the denoised image obtained with our algorithm described in Appendix~\ref{subappendix:pink}.
    Panels (d) and (e) show the output of the JWST pipeline algorithm (\texttt{clean\_flicker\_noise}).
    Panel (f) illustrates the difference between the two algorithms.
    Our algorithm effectively avoids over-correction near bright sources and better preserves diffuse emission.
    }
    \label{fig:pink_noise}
\end{figure*}

\subsection{1/f Noise Correction} \label{subappendix:pink}
The images taken with NIRCam are known to suffer from 1/f noise.
This is correlated read noise from the NIRCam detector readout system, caused by the readout integrated circuits and the direct current bias in the SIDECAR ASIC electronics \citep{Schlawin2020}.
An example of this noise is shown in Figure~\ref{fig:pink_noise}a, where horizontal and vertical stripe patterns are visible.
The horizontal stripes form four distinct bands corresponding to the detector readout structure, whereas the vertical stripes extend uniformly across the image.

To mitigate the 1/f noise, we first estimated the horizontal noise.
We divided the original 2048 by 2048 image into four subimages of size 2048 by 512, following the readout structure, and estimated the horizontal noise for each subimage.
Then we computed the derivatives along each vertical line.
Due to the correlated noise, the derivatives also exhibited characteristic stripe patterns corresponding to the horizontal noise structure.
We estimated the median derivative value at each row and subtracted it from the derivative images.
To reconstruct the image, we integrated the derivative field in the vertical direction using the first row of each original subimage as the reference.
Then, we combined the four horizontally denoised subimages.
We estimated the vertical noise with a similar procedure but without dividing the image into subimages.

Figure~\ref{fig:pink_noise}b shows the total 1/f noise for an example NIRCam exposure modeled in this work.
Figure~\ref{fig:pink_noise}c shows the denoised image obtained by subtracting the total noise from the original image.
This method allows us to measure the 1/f noise with much less influence from extended low SB features such as the ICL compared to measuring it directly from the original image.
For comparison, Figures~\ref{fig:pink_noise}d and \ref{fig:pink_noise}e show the 1/f noise estimate and the corresponding denoised image produced by the default JWST pipeline.
Figure~\ref{fig:pink_noise}f shows the difference between our result (Figure~\ref{fig:pink_noise}c) and the default JWST pipeline output (Figure~\ref{fig:pink_noise}e), revealing a non-negligible feature left by the default pipeline.
This motivates our customized correction, which more effectively suppresses the detector correlated stripes while minimizing contamination from extended low SB emission.

\subsection{Sky Gradients and Background Correction} \label{subappendix:sky}
Before mosaicking the final data, we corrected the sky gradient and background for each exposure. 
To measure the sky gradient, we adopted a linear plane model described by the following equation: 
\begin{equation} 
    I_{sky}(x,y) = a x + b y + c. 
\label{eq:model_gradient} 
\end{equation}
The coefficients $a$, $b$, and $c$ in Equation \ref{eq:model_gradient} represent the gradients along the $x$ and $y$ directions and the total level of the sky, respectively. 
The $x$ and $y$ values correspond to the pixel coordinates ranging from 0 to 2047. 
For each observation, we optimized the plane model to estimate $a$, $b$, and $c$. 
To suppress contamination from astronomical sources, we masked regions brighter than the 68th percentile of the overall pixel-value distribution. 
This threshold was chosen to systematically exclude pixels dominated by stars, galaxies, or diffuse emission, thereby minimizing bias in the estimation of the sky gradient. 
The fitted sky model was then subtracted from each image to remove large-scale linear gradients. 
Although this correction is small ($\mytilde0.1$\% at $\rm{29\ mag\ arcsec^{-2}}$), we applied the correction to ensure an accurate measurement of the diffuse ICL component. 

Next, we subtracted the residual sky background after mosaicking images.
To mask bright astronomical objects, we adopted the segmentation map generated by \texttt{SExtractor} as a base mask map.
Then we expanded the mask following Equation \ref{eq:expansion_width}.
We found that the estimated sky background converges when $c_{e}$ is greater than or equal to 2.0, which corresponds to a more conservative masking choice that prioritizes maximal contamination removal.
Therefore, we adopted the sky background measured at $c_{e}$ = 2.0 for the final correction.
Although the constant term c in Equation \ref{eq:model_gradient} already removes the first-order sky level, neglecting this additional correction introduces an error of about 0.03 percent at $\rm{29\ mag\ arcsec^{-2}}$.

We derived the SB limits $\mu_{lim}$ for each band by measuring the standard deviation of the sky pixel value distribution within a masked background region. This region was defined outside the 1 Mpc ellipse (SMA $\approx 119^{\prime\prime}$;  \textsection\ref{subsec:sbmeasure}) and covers a total area of  $\mytilde 1~\mathrm{arcmin}^2$.

To account for systematic uncertainties in the sky estimation $\sigma_{sky}$, we divided the background region into 20 azimuthal bins and measured the median sky value within each bin. We then adopted the standard deviation of these 20 mean values as the characteristic sky estimation error $\sigma_{sky}$. This systematic error is incorporated into the error propagation for all subsequent SB measurements.

\section{Cluster Member Selection} \label{appendix:member}
\begin{figure}
    \centering
    \begin{minipage}[t]{\columnwidth}
        \centering
        \includegraphics[width=\linewidth]{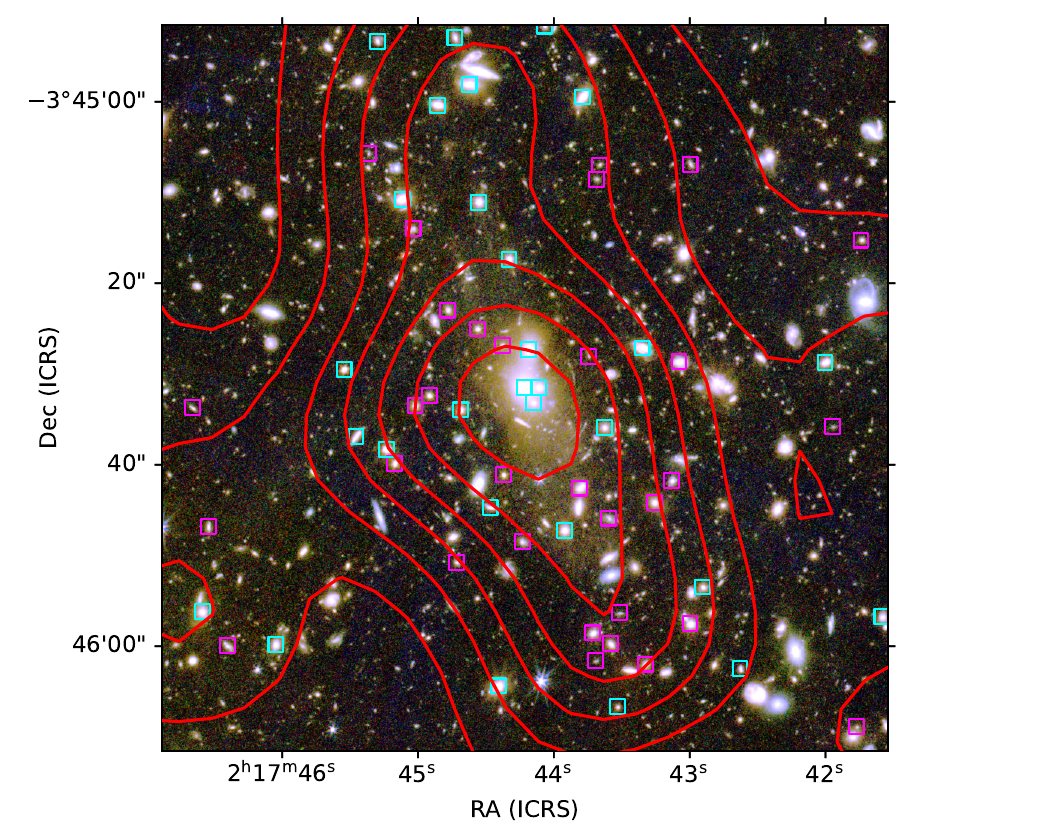}
    \end{minipage}
    \hfill
    \begin{minipage}[t]{\columnwidth}
        \centering
        \includegraphics[width=\linewidth]{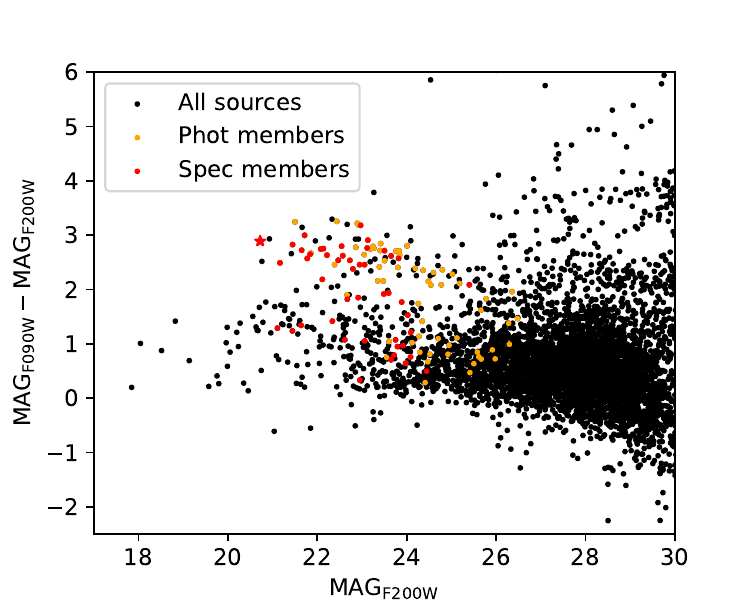}
    \end{minipage}    
    \caption{
    (Top) Color-composite image of XLSSC~122 with red for F356W, green for F277W, and blue for F090W+F200W.
    The cyan boxes indicate spectroscopically confirmed member galaxies and magenta boxes show photometric members.
    The red contours show the smoothed number density distribution of all known member galaxies.
    (Bottom) Color-magnitude diagram of the JWST XLSSC~122 field.
    Black dots indicate all detected sources.
    Red and orange dots show spectroscopically and photometrically selected member galaxies, respectively.
    The star represents the BCG.
    }
    \label{fig:01_colormap}
\end{figure}

Previous spectroscopic analyses identified 50 cluster member galaxies in the XLSSC~122 field \citep{Willis2020}.
We now have seven-band photometry from HST and JWST, which can aid in cluster member identification further. Photometric measurements were performed in the dual-image mode of \texttt{SExtractor} \citep{Bertin1996}, where a detection image was created by combining the four JWST filters, and fluxes were measured in all seven bands.
Photometric redshifts were then estimated with \texttt{eazy-py} \citep{Brammer2008}.

Since the cluster lies at $z\sim2$, distinguishing background galaxies from cluster members becomes non-trivial.
At these redshifts, photometric redshift uncertainties are relatively high due to the limited wavelength coverage of the key spectral features and the degeneracy between redshift, dust extinction, and stellar population age.
Moreover, the SED of high-redshift galaxies can differ from those of low-redshift counterparts, as they are often dominated by younger stellar populations, higher star-formation rates, and stronger dust attenuation.
To improve the robustness of our member identification, we adopted the high redshift optimized SED templates from \citet{Larson2023}, which are optimized for high redshift photometric redshift fitting and can reduce low redshift interloper solutions, and may help mitigate contamination in member selection.

Using the photometric redshift of each source and its associated uncertainty, we selected only those with uncertainties smaller than 25\%.
The range of selected member candidates lies at $1.7\lesssim z_{phot} \lesssim 2.3$.
We then used the spectroscopic catalog of \citet{Willis2020} to remove spectroscopically confirmed foreground objects and to retain 50 confirmed cluster members.
After this spectroscopic cleaning, our photometric selection contains 90 member candidates, of which 60 are newly identified in this work.
For the 30 sources with both photometric and spectroscopic redshifts, we find good agreement, with $|z_{phot}-z_{spec}|/(1+z_{spec})<0.1$.

In the top panel of Figure~\ref{fig:01_colormap}, cyan boxes mark spectroscopically confirmed members, while magenta boxes denote photometric members.
The distribution of member galaxies is centered around the BCG and shows an elongated morphology along the north–south direction, with a pronounced extension toward the south.
This elongation is aligned with the major axis of the BCG and the ICL in XLSSC~122 measured in \textsection~\ref{subsec:sbmeasure}.
As shown in the bottom panel of Figure~\ref{fig:01_colormap}, the distribution of these members in the color–magnitude diagram exhibits a well-defined red sequence.
In addition to the red members, we also identified a blue population, consistent with ongoing star formation activity.
These photometric member distributions in the bottom panel of Figure~\ref{fig:01_colormap} are consistent with those found in \citet{Willis2020}, and they extend to sources up to two magnitudes fainter than the spectroscopic members.

\end{appendix}

\end{document}